# Bio-mimic four-dimensional printing of nanostructured interactive hydrogels

*By Yufeng Tao*

Yufeng Tao  wsnwp520@sina.com

Chengchangfeng Lu
Whiting School of Engineering, Johns Hopkins University, Baltimore, 21218-2688 (USA)

Nature has exhibited numerous soft-interactive botanical or zoological systems with self-adaptivenss or reconfigurable shape, however, artificial micro/nano materials of same merit are formidable due to material/structural limitations, posing multidimensional challenges to current chemical/physical fabrications. Here, we propose a new nanostructured hydrogel by four-dimensional (4D) printing for function integration and dimension reduction. The cross-linked matrix relaxes or contracts in robust hierarchical heterojunction to eliminate interface issues of dual-layer desgin. Nanowire operates as minimum re-programmable unit (the smallest stimuli-responsive hydrogel, volume less than one fiftieth of standard kirigami lattice). 4D hydrogels promise solvent/light programmability, ultrafine feature size (<220 nm), large bending angle (>1000°), variable Youngs' modulus (>20 Mpa) and self-recovery ability. We fabricated several homogeneous in/out-plane grippers, bio-mimic hands/muscles/actuators, soft scaffolds, buckled microbridges and photon filters as verfication. Taking use of this maskless 4D printing, many more programmable robotics/photonics devices far beyond demonstrations here could be created.

## 1. Introduction

Four-dimensional (4D) printing renders permanent structures with time-participated shape programmability, [1] opening up a new scientific frontier with unimaginable potential in material design, prototyping, manufacturing and assembly.[2] The shape-morphable hydrogels are one of pioneering 4D products (artificial grippers/robotics,[3] deployable cages/scaffolds,[4] drug delivery, [5] tissue regeneration, [6] optics/photonics, [7] soft machines, [8] and strain sensor).[9] However, current-stage hydrogels have yet to achieve perfection due to isotropic swelling or shrinkage, [6, 7, 10] slow water diffusivity ($<10^{-9}$ $m^2s^{-1}$), [11] inadequate stiffness and poor function. Microcosmic environments (the blood vessel, vein, artery, skin pore, microfluidics or on-chip labs) are physically inaccessible for macro hydrogels.[3-11] Thus, reducing dimension of functional hydrogel become a prerequisite for integrating robotic applications. [12,13]

The last decade witnesses an explosive application of two-photon polymerization (TPP) across a broad range of mechanics/bio-science/optics devices. [14,15] This predominant real-3D microstructuring tool [16] generally produces inanimate sculptures deprived of time-depending behavior. [17] Therefore, researchers often fabricated compound bilayer/trilayer structures for controllable stimuli-responsiveness. [18,19] Unfortunately, bilayers require tedious multi-step multi-material (at least active and inert materials) process, often transit one-way if releasing out residual stress, and suffer from poor interfacial dis-matching.[18,20-22] Till now, multi-functional hydrogels *via* homogenous material in rapid manner remains challenging but highly-valued.[21]

In nature, plants/animals utilizes water sorption-to-desorption for directional motion or shape morphing.[23] Living organisms of hard-soft arranged hierarchical architectures wriggle to transport in/out substances. Pinecones self-open/close according to drying/humidifying. Sea cucumbers stiffen or soften by vomiting/taking water. Mimosa shut by directional movement of water molecules on upper and lower surface of leaf pillow. Tendrils and bracts varies internal turgor by applied stimuli. Natural materials such as fish scales, spider silk or nacre demonstrated outstanding shape-changing properties for survival under external shock. Butterfly wings or chameleon skin reflected multiplex structural colors. [24]

To overcome technological bottlenecks and integrate bio-mimic functions on-chip, [25] we propose a covalently-interconnected heterojunction to TPP. Amphiphilic polyethylene glycol diacrylate (PEG-DA) [26] was photon-incorporated with N-isopropylacrylamide (NIPAM) [10] at high-order path modulation, obtaining mechanics-tunable nanostructured interactive hydrogels (MNIHs, **Figures 1**, S1-S3, Video 1). Following that, we demonstrated in/out-plane interactive grippers, tunable photon filters, bio-mimic hands/muscles (Figure 1a), light-fuelled actuators showcasing some tempting benefits to science commnities:

1. The nanowire (NW, linewidth <100 nm, height <800 nm if using 100×oiled objective) operated as minimum programming unit (the smallest hydrogel, at least two-magnitude smaller than standard origami), guaranteeing ultrafine resolution in spatiotemporal reconfigurations.[27]

2. Self-driven mechanism was a task taken by material scicence for pushing frontier of robotics. Here, intermolecular hydrogen bonding affected net-point number, making MNIHs self-control shape without organs/generators or sensing components.

3. MNIHs contained foldable volume of loosely-linked intact network (Figure 1b) for self-recovery if undergoing unwanted deformation, which improved longevity of devices, and decreased repairing cost/complexity.

4. Rapid fabrication provided homogenous hydrogels, no interaction issue existed like dual-layer devices, MNIHs maintained desirable structural stability in reverse reconfiguration.

More merits such as high fatigue resistance, large force-to-weight ratio, predictable high-freedom programmability (Figure 1c) and desirable optical clarity could be found in this work.

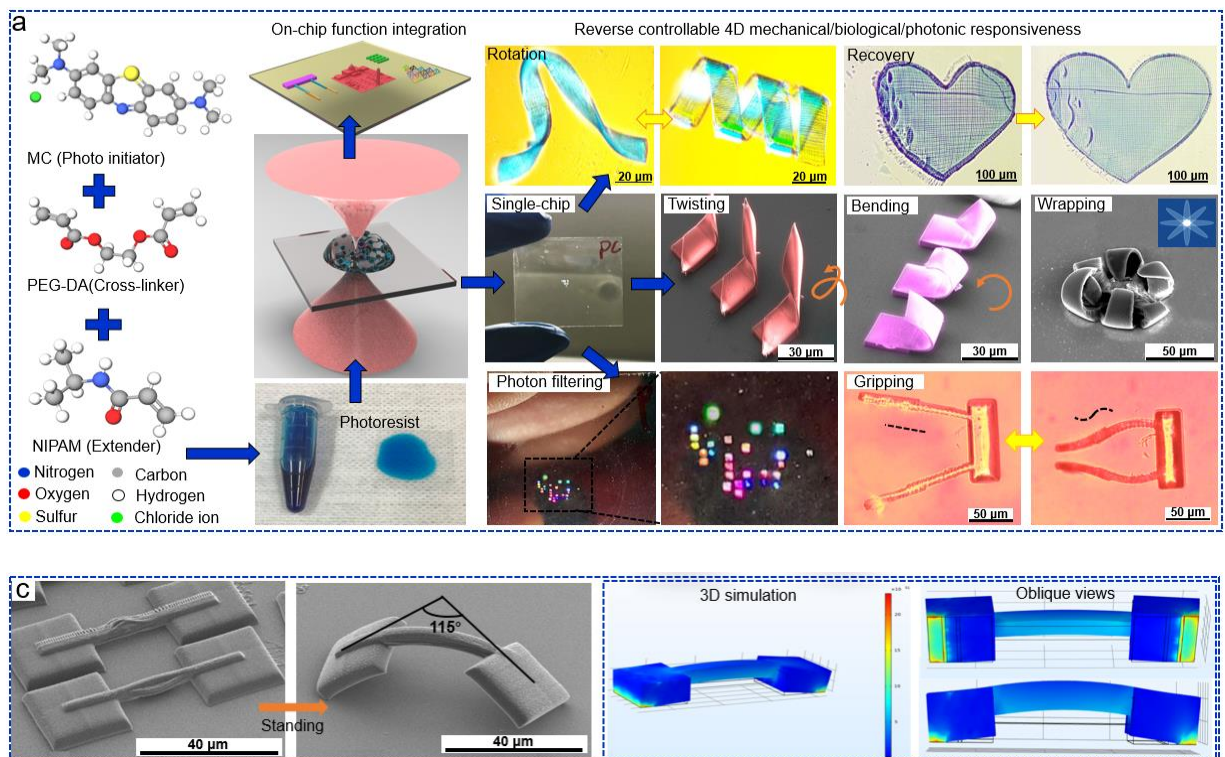

**Figure 1.** (a) Schematic of femtosecond laser processing interactive materials for devising multi-functional MNIHs. (b) Experiment optical system and laser manipulating NWs to control intermolecular force and mechanical compliance. Length, direction and density of NWs jointly determine shape programming. (c) SEM images of a cantilever self-standing on substrate by shape programmability and qualitative morphological computation by finite element modelling to predict morphing trend and stress distribution.

## 2. Results

Cationic phenothiazine, methylthionine chloride (MC) contained nitro and sulfur groups as electron acceptors for higher-probability photon absorption (doping concentration as low as 0.5 wt%). [28,29] The prepared photoresist demonstrated suitable adhesion/stress/mobility on substrates during single-step fabrication of various 3D architectures (Figures 1a, S1 and Video 1). Detailed chemical reaction was analyzed by Fourier transform infrared spectroscopy (FT-IR, seen in Figures S4). [30] Concomitant photon-excited fluorescence highlighted near-infrared beam for positioning or high-fidelity imaging. Laser travelling route implanted MNIHs with initial geometric shape memory (Figure S1). MNIHs converted osmotic chemicals to kinetic energy, [31] where functional groups of PEG-DA/NIPAM activate intact network to perceive external stimuli or desorb matters to introduce nonlinear mechanical behaviors (Video 1). [32]

*Transformative helix*

Helix (or "spirochete") denoted a typical structure in robotic swimmers or holders. [1-4,12,19] Here, we fistly constructed 3D "bracelet" (**Figure 2**) at a horizontal expansion > 120% and a lateral expansion < 8% from a planar (dimensions of 30×150×4 μm$^3$, scanning speed of 1 mm/s, costing 50 sec). Then, we intentionally-titling NWs to steer bending direction, MNIH self-buckled, folded, rolled and rotated forcefully at least three turns ($\beta$ > 1000°, tilting angle equaled 30º, Figures 1b and 2b), forming an out-plane DNA-like "spirochete", or shrunk, opened, lied down to starter shape (Video 2). No any breaking-up or collapsing was found in repeated reconfiguration.

By carefully determining position of NWs, the shape morphing could occur throughout the entire hydrogel, or just locally occurred in specific areas. Multiple 3D forms self-established from planar starting shape, triggered by single droplet of solvents/oils. We regulated local duty cycle for multi-point buckling (Figure S5). Individual "arch bridges" spontaneously stood at specific coordinates, regions of bigger $\Delta L$ buckled out-plane as cross-sectional view evidenced by scanning laser confocal microscopy (SLCM, Figure S5). Experiment results revealed the bending angle $\beta$ depended on trade-off balanced by line pitch $\Delta L$ and NWs' width $\Delta N$.

In cycles of interactive transformation, cross-linked network ensured tunable mechanical properties and desirable structrual integrity during nonlinear reconfiguraiton. [33, 34] A small deviation on rotating angle of less than 5% (Figure 2b) satified most of actuation requirements. The nanostructured heterjunction exhibited highly-directional bending, twisting, rotating and buckling-up with desirable repeatability (no observable fatigue phenomenon after 1K time transformation), which took less responsive time than macro hydrogels, accelerated device design-to-prototype process, and allowed multiple geometries from same starter shape.

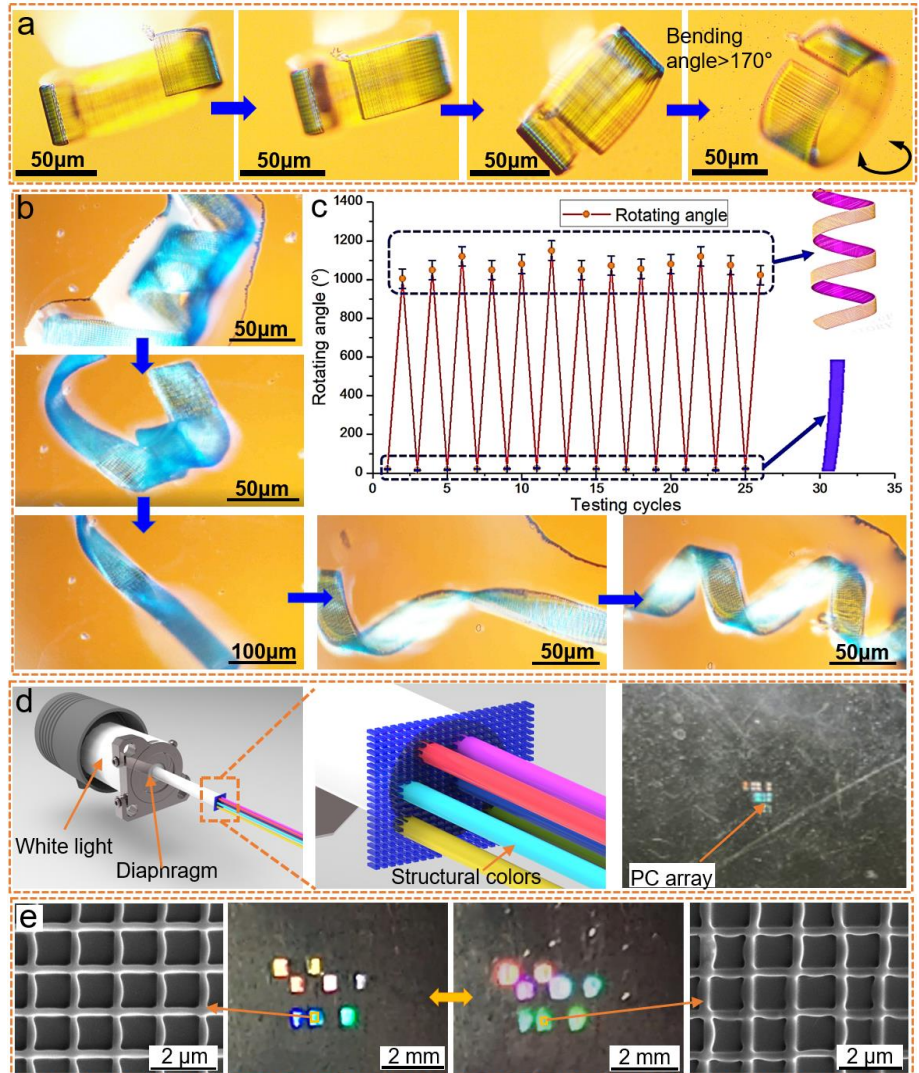

**Figure 2.** (a) Accumulation of intermolecular force realized pre-designed responsiveness. Constant-ΔL MNIH transformed to a "C" bracelet. (b) Titling-NW MNIH crimped into 3D helix at three turns reversely. (c) Repeated transformation confirmed a large bendable angle exceeding 1080°. (d) Experiment setup for on-chip generating structural coloration from wide-spectrum white light source, and on-chip PC arrays. (e) Shape-deformable PC array, brilliant coloration red-shifted and broadened through self-assembly, zoomed-in SEM images of NWs before and after swelling.

*Reconfigurable photonic crystal*

4D-printed hydrogel nanowires synergized shape-morphing capability and high-resolution nanostructure for nano-photonics (photonic crystal, PC), allowing PC to dynamically effect optical field.[7] We developed on-chip interactive photonic crystals consisted of NWs to filter out different-frequency photons as chip-integrated light sources, memories or displaying. [24] A full-color chip reflected broad-band structural colors (purple, blue, green, yellow, orange, red and their transitional-wavelength color in Figures 1a and 2d) by carefully modifying the scanning path and optics parameters (Figure S7).

Temperature/solvents transformed periodic structural distribution and dielectric constants to modulate photonic bandgap, [35] which shifted and broadened optical frequency for photonic engineering. Exampled by on-chip generation of coloration *via* single-layer PC as photon filters (size of 240×240 $\mu m^2$ for each), wavelength-controllable highlighted structural colors irradiated from a small area of single chip. PCs could be immersed in water to hide information, or heated/ humidified to show structural colors filtered from a white-light source (Figure 2d), as dynamic labels used in encode information and anti-counterfeiting. [24, 35] Self-assembling PEG-400 modified NWs' morphology and changed their dielectric constant difference, therefore, structural coloration red-shifted and broadened (Figure 2e). Using our proposed interactive material formula and 4D printing techniques, all researchers could easily design, tuning and modelling their dynamic hydrogel photonics for numerous programmable optical components (grating, modulator, coupler or display plane) of simultaneous shape/light manipulation, or meta photonics. [35]

*Bio-mimic hand and grippers*

Reusable high-freedom holding or gripping was fundamental manipulation in microfluidic, robotics or bio-sciences. [2-5, 25] MNIHs promised various in/out-plane muscle behaviors for satisfying this aim (**Figure 3**). Exampled by a five-fingered dexterous palm precursor (Figure 3a), its thumb bent upward, and index/middle/ring/little fingers gradually clenched into a fist. Similarly, closure of apricot flower-mimic MNIH (Figure 3b) and a cross-pattern gripper (Figure 3d) were programmed, we theoretically-reconstruct the shape reconfigurations in the frame of an explicit finite element method (Video 3).

Far beyond simply programming shape-changing direction, bending shape and triggering condition of arms became programmable by arranging different heterojunctions. Here, an in-plane gripping MNIH was designed to change its C-arm to S-arm (Figures 3c and S6, Video 3), simultaneously, the inspiring condition switched oppositely. Water made C-arm closed, but S-arm opened. All in/out-plane grippers self-opened or closed (Figure S6) in absence of built-in electric/pneumatic drivers for implanting robots of different situations. [12, 25] We programmed a swellable woodpile (dimension of 80×80×30 μm$^3$, contracted in air and expanded by diluted blood as cage machines, potentially assisting to trap cells/microorganism, Figure S6, Video 4 ).

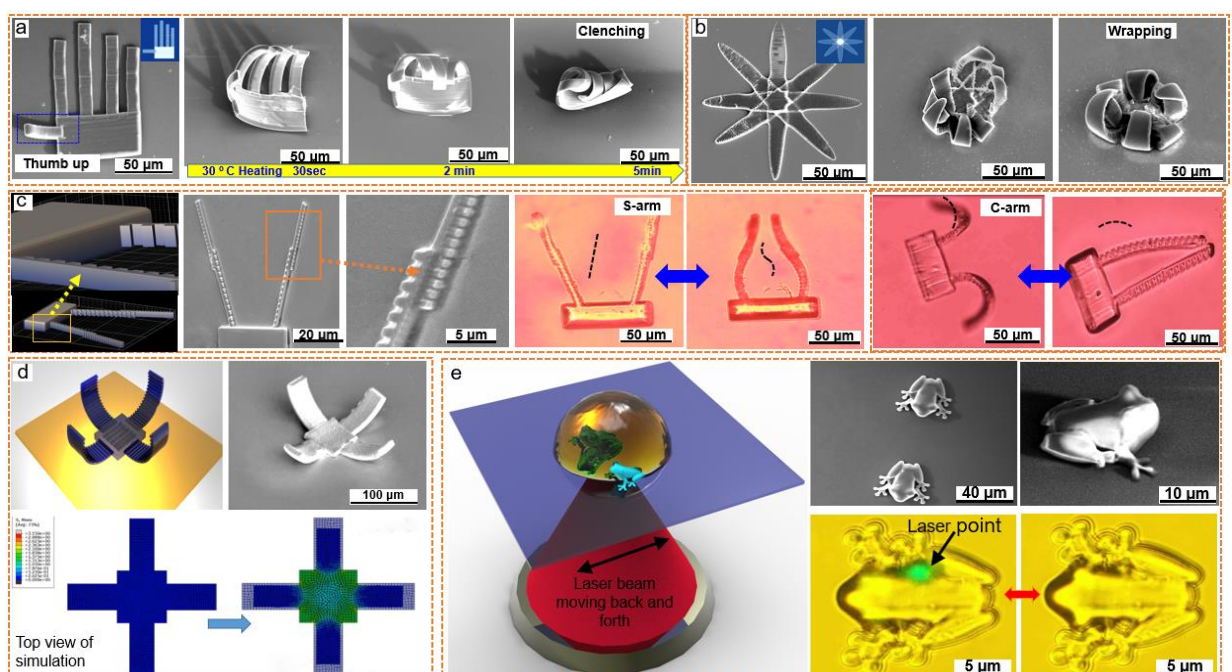

**Figure 3.** Harnessing the temperature-triggered shape deformation provided great interest to emerging micro-mechanic/bio-science fields and was identified here. (a) A hand-mimic MNIH gradually curled into a clenched fist. (b) A heat-triggered flower-mimic closure. (c) Models and SEM images of in-plane S-shape arm of soft-interactive grippers, and the traditional C-shape arm. (d) A cross-shaped out-plane gripper and top view of finite element simulation. (e) A frog-mimic MNIH in water for non-contact laser driven actuation, SEM images, and actuated shape change due to laser point.

*Laser-driven actuator*

Outperforming local/entire shape programmability, MNIHs supported more useful 4D behavior, the widely-concerned reverse light-deformable ability.[36] The unique NIPAM/PEG-DA network possessed intrinsic photon-thermal conversion ratio without doping single-walled carbon nanotube or graphene. Significantly different to solvent diffusion process, light-driven thermo-responsive behaviors replied on no solvent/chemicals. [37,38] The laser-projected region of MNIH generated local stress variation and expanded for pointed shape morphing, therefore, an external laser can remotely trigger micro-scale actuation.

A free-standing frog-mimic MNIH was fabricated by 4D nano printing and put in water for swelling (Figure 3e). Subsequently, we laser-scanned this fog externally, the photon-thermal conversion redistribute the local stress inside MNIHs, therefore, frog became living and nodded tirelessly (Video 5). When laser-point was remove, frog recover to stationary gesture. Long-term results confirmed that nodding frequency and amplitude strongly depended on laser scanning speed and optical power, no fatigue was found after at least 10000 times actuation.

*Self recovery*

Another competitive advantage, MNIHs promised intrinsic impact/shock-absorbing ability outperforming permanent sculptures (covalent bond-formed structures broke into waste if deformed). [2,3] Even being heated, distorted or squeezed by externally-applied force, MNIHs not disintegrated but self-recovered to shape memory by absorbing organic matters (< 0.1 mL, **Figure 4**), implying a remarkable survival ability in extreme conditions. We *in-situ* fixed a "broken heart" (dimension of 500×700×15 μm$^3$) distorted by a spun metal tweezer beforehand to pre-designed symmetrical shape (recover rate exceeded 90%, Video 6), the cavitation-induced air bubbles escaped away.

Mobile hydrophilic supra-molecules worked as nano-binders for regeneration in defected morphologies. [9] Strength of hydrogen bonding depended on surface energy, interfacial condition, electrostatic attraction force and volatility of materials. [9,15] Following that, we poked a woodpile MNIH using a sharp needle, leaving a collapsed hole, then dropping PEG-400 as nano-binder, woodpile absorbed PEG-400 molecules and apparently repaired holes (Figure 4c, Video 6). Mechanical testing shown the repaired hole region reached over 70% Young's modulus (12MPa) of normal state (15 MPa, Figure 4d), and initial feature get recovered.

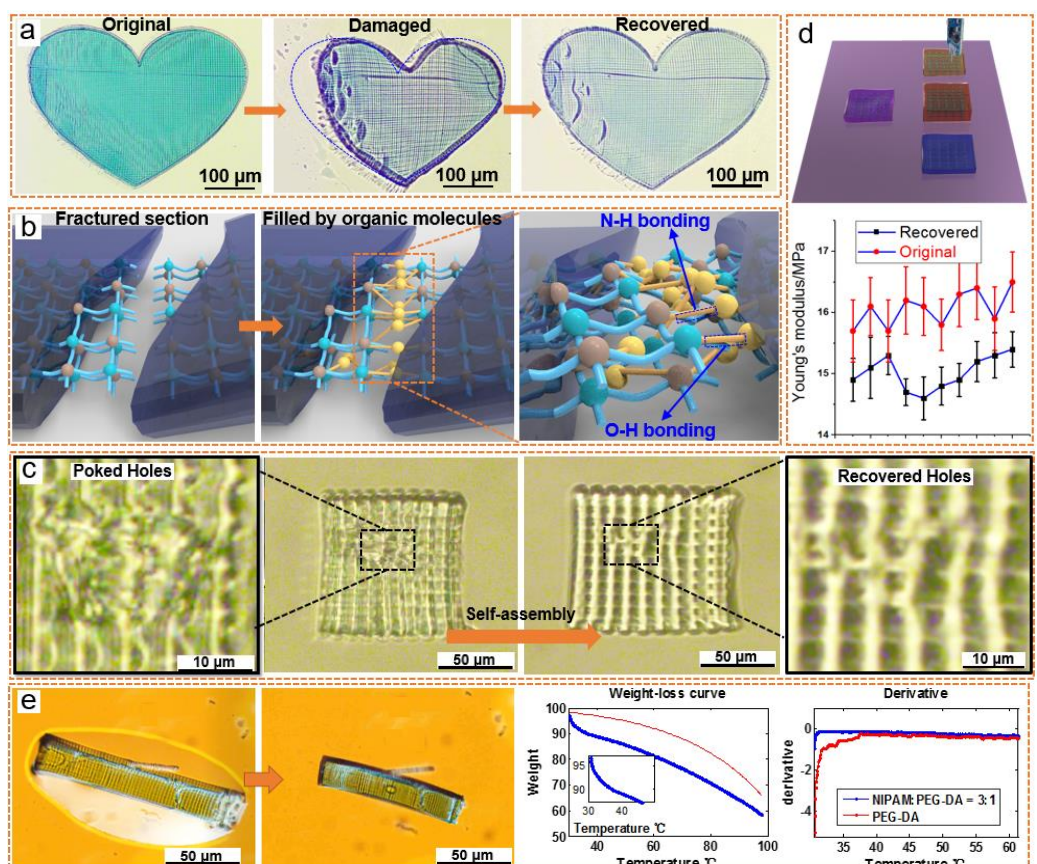

**Figure 4.** MNIHs got repairing by self-assembly according to shape memory. (a) An squeezed malformed "heart" self-recovered to initial symmetric heart. (b) Illustration of self-assembling molecules in fissures. (c) A needle-poked hole self-reconstructed by absorbing PEG-400, which exhibited a new on-line repairing method for on-chip functional devices. (d) Schematic of mechanical test, and Youngs' modulus of initial and recovered MNIHs. (e) Shrinkage of MNIHs (Video 7), weight-loss curve and its derivative in TGA test. Cross-linked hydrogel (blue color) presents substantial weight loss around $LCST_{NIPAM}$. Pure PEG-DA-based, polymeric-tested sample showed little temperature sensitivity within a zone of 30-45 °C.

*Solvent retention*

Self assembly of hydrogel was mirrored by solvent retention among MNIHs.[39] Reswelled MNIHs were taken thermogravimetric analysis (TGA) for analysis solvent retention ability.

Results shown relative weight loss varied differently with or without extender NIPAM (43% and 36%, respectively Figure 4e). NIPAM monomer transited hydrophilicity/hydrophobicity at lower critical solution temperature ($LCST_{NIPAM}$=32.4 °C), and PEG-DA presented monotonous hydrophility below 100 °C (Figure S2). 4D nano-printing changed density and thermal conductivity of chain-style polymerized PNIPAM, making hydrophobic behavior occurred within a broadened 30~42 °C zone. Seen in measured results, the hybridized NIPAM/PEG-DA matrix decreased over 8% weight around $LCST_{NIPAM}$, while pure PEG-DA smoothly lost only 2.1% (Figure 4e).

*Nanoscale morphologies*

To identify nanoscale morphologies of swelling and shrinkage, we post-processed MNIHs through freezing-to-drying for nanometric morphologies, SEM images presented an interlaced, folded, nanofibrous-like long-chain conformation rather than porous morphology of pure PNIPAM (**Figure 5**), and sustained a big interfacial contact area for mobility improvement. Local molecular conformation switched for folding and unfolding states (Figure 5a), which effected the density/mechanics of MNIHs significantly.

*Mechanical properties*

Superior virtue of both stiff and soft mechanical property was pivotal to address challenges of biological interconnects, biomedical devices, soft machines/robots and facility.[34,40] Young's modulus (E) of stimuli-responsive MNIHs exhibited a wide MPa variation for generating tensile force. We conducted an in situ force-sensing probe of a micromechanics platform to penetrate into (compressive) and pull out (tensile) from MNIHs for stiffness (Figure 5b). Synchronous stress (load force) and displacement were recorded and data-fitted for E (Figure 5c), an elastic curve represented a hysteresis loop, confirming MNIHs as mechanics-tunable material.[40]

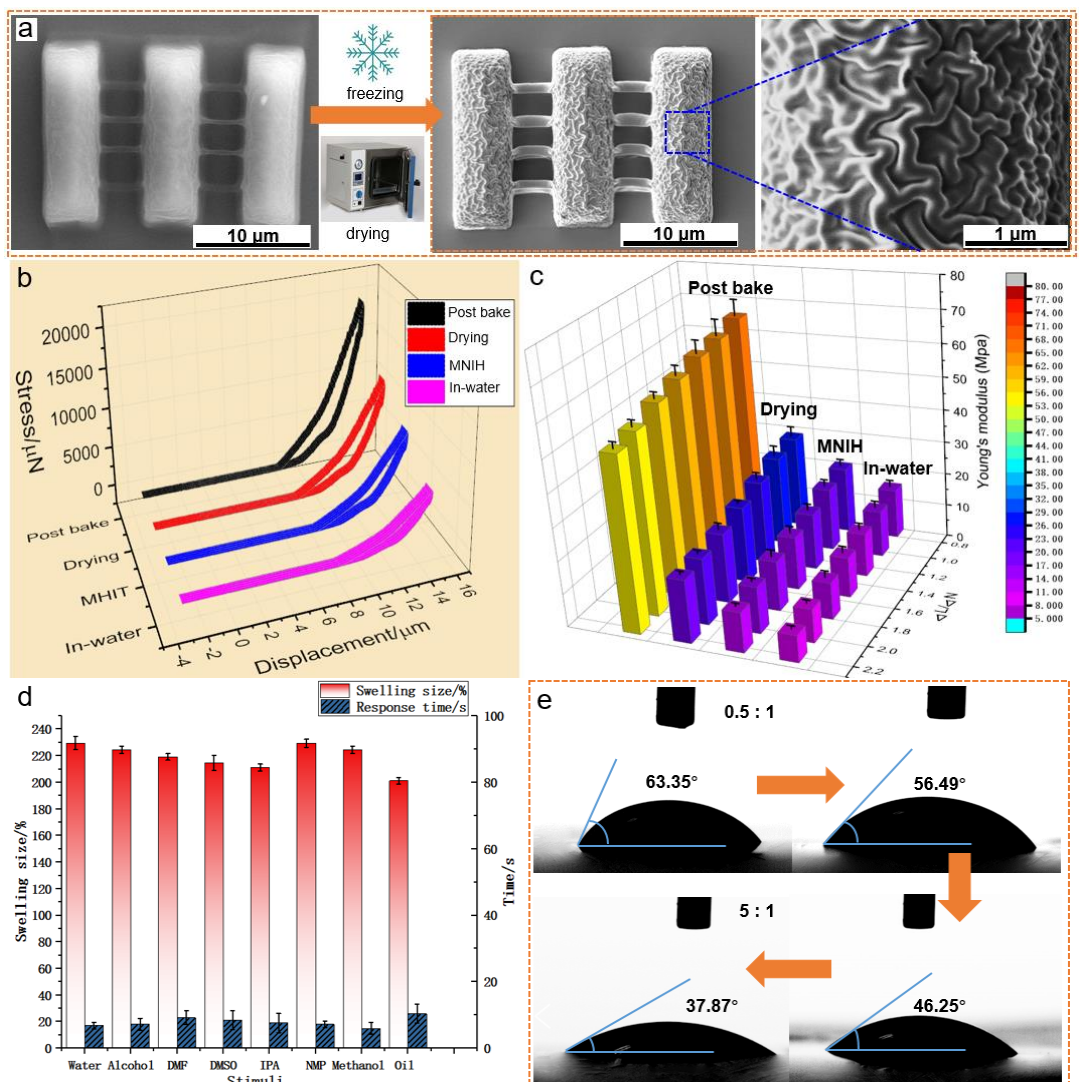

**Figure 5**. (a) Cantilever MNIH was equilibrium-swelled, subsequently freezed at -5 ºC for 4 h and vacuum-dried (DZF-6020, Boxun) at 40 ºC for SEM characterization. (b) Hysteresis loop of stress-to-displacement data (stiffness) in loading and un-loading processes, where post-bake, drying, MNIH, and in-water refer to the 100 ºC baked MNIH, 30 ºC-dehydrated MNIH, the just-developed MNIH, and the water-immersed MNIH. (c) The summarized Young's modulus of MNIHs at $\Delta L/\Delta N$ (Spacing/NW width) ratio. (d) Different swelling performance of available osmotic organic matters (DMF, DMSO, IPA and NMP refer to dimethylformamide, dimethyl sulfoxide, isopropanol, and N-methyl pyrrolidone respectively). (e) Dynamic water contact angle measured on surface of different material ratio MNIHs.

MNIHs achieved a standard tunable MPa level $E_o$ of $18 \pm 6$ MPa (by changing spacing and laser exposure dose, where the Poisson's ratio $v = 0.49$), [41] which decreased to 4~13 MPa immersed in water, increased to 21~34 MPa if evaporation occurred, and jumped to $65 \pm 8$ MPa after 100 ºC post-bake (Figure 5c), which was stiffer than elastomer poly dimethylsiloxane (PDMS) ($E_{PDMS}$ was 5~20 MPa) or Ecoflex™, but softer than poly methyl methacrylate (PMMA) ($E_{PMMA} > 1$ GPa).

*Stimuli materials*

In addition of the existing stimuli (temperature, pH, light, ion and water), our experiment confirmed more available osmotic organic matters (dimethylformamide, dimethyl sulfoxide, isopropanol, N-methylpyrrolidone, styrene sulfonate, methanol and glycol, Figure 5d), tissue liquids (sweat, eye tears, blood, saliva and even urine) and even edible oils can drive MNIHs out of equilibrium for shape reconfiguration. Acetone-type solvents also can trigger MNIHs in a few of time but with destructive solubility.

MNIH automatically investigate stimuli without sensing components (in stark contrast to sensor-embedded robot), which would escalate the interest of sensory materials. Huge impact on responsivity could be created by altering photoresist formula, exposure dose or $\Delta L$ created a huge impact on responsivity (Figure S), as summarized, water-contact angle (Figure 5e) declined 63.35°→56.49°→46.25°→37.87° on order of NIPAM:PEG-DA ratio increased 0.5:1→1:1→3:1→5:1. Hydrophilic/oleophylic MNIHs self-assembly organic solvents/oils for organized deployment, potentially revolutionize therapeutic and diagnostic procedure triggered by tissue liquids that was not possible otherwise.

*Self-driven mechanism*

Trick of the new self-driven mechanism lied on manipulating functional groups density to accumulate intermolecular-based reconfigurable stress. As bi-functional material, MNIHs generated negative angular momentum in swelling, driving them to an "*n*" or "*c*" shape, however, thermal shrinkage generated positive angular momentum to lift upward as a "*u*" or "L" shape. If the induced tension defeated adhesive constraint, out-plane actuation occurred. Otherwise, MNIHs underwent in-plane shape morphing. To empirically explain shape reconfiguration, a theoretical model delineated the relationship between the Young's modulus $E$ and surface tensile force $\Delta F$. Seen in Equation 2 ($I$ = moment of inertia, unit: Kg·m$^2$; $\rho$ = radius of curvature in the beam deflection theory, unit: µm$^{-1}$), declining -$\Delta E$ created a negative bending moment $\Delta M$ (unit: N·m) over the rectangular hydrogel sheet. Surface tensile force $\Delta F$ resulted from integration calculation of local residual stress, $\Delta\sigma$, (unit:N) on single-layer thickness $H$ through entire length $L$ (Equations 3 and 4, $\varepsilon$ = local strain, $v$ = Poisson ratio of the fabricated gelation).

$$\Delta L = k\Delta N, k \in [0.5\ 2.5] \quad (1)$$
$$\Delta M = -\Delta E I/\rho \quad (2)$$
$$\Delta \sigma = -\Delta E \varepsilon/(1-v) \quad (3)$$
$$\Delta F = \int_0^H L\Delta\sigma dz \quad (4)$$

Dynamic $E_m$ was data fitted by averaging stiffness of $\Delta S/\Delta D$ in penetrating and pulling-out processes as Equation 5:

$$E_m = 1/2\, m\, P_w\, T_t(\Delta S_+/\Delta D_+ + \Delta S_-/\Delta D_-). \quad (5)$$

where $m$, $P_w$, $T_t$ denoted coefficients of stiffness-to-Young's modulus, effective laser power and interaction time, bending angles followed below exponential relationship (Equations 6 and 7). Angle was $\beta(t)$, $t$ denoted time, $\tau$ was a time constant determined by material, and $L$ denotes the lateral bending length. The observed curvature, $\kappa$, reached 1 µm$^{-1}$ (Figure S6e).

$$\beta(t) = \beta_o[1 - exp(t/\tau)] \quad (6)$$
$$k = \beta/L \quad (7)$$

### 3. Conclusion

4D nano printing cooperated with interactive composite materials to unleash capacity of hydrogels breaking constraints of traditional TPP, which reduced dimension, shortened design-to-prototype period, improved shape programmability. Functional groups governed self-assembly for local or throughout softness-to-stiffness change. MNIHs self-interact with osmotic chemicals, light or temperature to store/release mechanical energy, where cross-linked long-chain conformation worked as neuromorphic matrix. Here resolution was selective (Figure S7), defect was self-repairable, mechanic properties were reversible tunable (4~30 MPa), time and material cost was saved. Moreover, intergrating individual micro-machines/nano-photonic/bionic device by our proposed method promised an unimaginable potential of holistic systematic applications.

## 4. Experimental Section

*Material preparation*

All ingredients were available from Sigma-Aldrich. We prepared glycol (($CH_2OH)_2$, 0.3 mL, 99+% purity, molecular weight Mw = 62.068, solvent), NIPAM ($C_6H_{11}NO$, 0.6696 mg, 99.99+% purity, Mw = 113.16, extender), PEG-DA (($C_3H_3O).(C_2H_4O)n.(C_3H_3O)$, 1.8 ml, 99.999+% purity, Mw = 400-700, cross-linker) and methylthionline chloride (MC, $C_{16}H_{18}ClN_3S$, 99+% purity, photon initiator). The mixture of PEG-DA/NIPAM was added to potassium persulfate ($K_2S_2O_8$, 0.001 mg, Mw = 270.32) for pre-polymerization at room temperature. Subsequently, more NIPAM (0.3348 mg), PEG-DA (0.9 mL) and MC (0.03 mg) joined the pre-polymerized mixture under magnetic stirring at 800 rpm for 1h. Photoresist was centrifugation purified by removing sediment before usage (Figures S2 and S3).

*Morphology characterization*

Substrates were pre-coated with conductive indium tin oxide (ITO) (100 nm thickness) film for electric conductivity. Nanoscale characterization replied on the field-emission electron microscope (FEI Nova NanoSEM™ 450) at acceleration voltages of 2-10 kV, magnifying 100-30,000 times, or an ESEM (Thermo Scientific™ Quanta™ 200, FEI) on low-vacuum condition for secondary-electron images. All MNIHs were freezing-to-drying processed before SEM for more details. We also utilized scanning laser confocal microscope (LEXT OLS5000™, Olympus) to nondestructively evaluate surface profile at subwavelength accuracy.

*Laser direct writing*

A mode-locked femtosecond Ti:Sapphire laser (Chameleon Discovery, Coherent) emitted a wavelength-tunable, pulsed laser beam (80 MHz repetition rate, 100 fs pulse width) to initiate a cross-linking reaction (power density of 2 to 20 mW/μm$^{-2}$, exposure time of 0.4 to 8 ms). Available laser wavelength covered visible and invisible ranges (754 to 956 nm). A terminal microscope contained a charge-coupled device (CCD), dichroic mirror, and objective-selective system. A close-looped, nano-step piezo displacement stage (300 × 300 × 300 μm$^3$ range and 0.2 nm step size) moved in predesigned 3D trajectories (Figures S1 and S7).

*Thermogravimetric analysis*

We conducted thermogravimetric analysis (TGA) to real-time monitor changeable solvent retention versus temperature. Equilibrium-swelled MNIHs underwent water evaporation when ambient temperature ramped up at 5 °C/min by a thermogravimetric analyzer (Q500, TA Instruments, resolution of 0.1 μg) within low temperature zone (30 ~ 100 °C).

*Optical spectroscopy/imaging*

To elucidate photon-chemical reaction during hydrogelation, Fourier transform infrared spectroscopy of powders (NIPAM, MC), liquid (PEG-DA), and MNIHs were performed by a Fourier Transform Infrared Spectrometer (FT-IR) (Nicolet™ Nexus 670, Thermo Scientific™) at wavenumbers of 500 to 4000 cm$^{-1}$ to deduct molecular structural transferring (Figure S4).

*In situ micromechanical measurements*

MNIH possessed magnificent mechanical programming properties that could be harnessed for dynamic applications. Herein, an advanced micromechanics testing instrument (FT-MTA02, FemtoTools) was deployed to investigate MNIH's stiffness, cohesive behavior and output force amplitude. The tungsten probes (2 μm tip radius) contained an in-packaged capacitive force sensor to reflect mechanics during cycles of loading/unlading process at 5 nN resolution in 1000 μF range. Heating operation was realized by a thermoelectric cooler at 0.625 °C resolution within -20~80 °C range or galvanometric laser-scanned MNIHs for photo-thermol conversation.

**Supporting Information**
Supporting Information is available from the Wiley Online Library or from the author.

**Acknowledgements**

This research was financially supported by the National Key R&D Program of China (2017YFB1104300), National Natural Science Foundation of China (61774067), National Science Foundation (CMMI 1265122), National Science Youth Fund of China (61805094), and the Fundamental Research Funds for the Central Universities (HUST:2018KFYXKJC027).

Received: ((will be filled in by the editorial staff))
Revised: ((will be filled in by the editorial staff))
Published online: ((will be filled in by the editorial staff))

**References**

[1] F. Momeni, S. Mehdi Hassani.N, X. Liu, J. Ni. A review of 4D printing, *Mater. Design.* 2017, **122**, 42–79

[2] S.D. Miao, N. Castro, M. Nowicki, L. Xia, H.T. Cui, X. Zhou, W. Zhu, S. Lee, K. Sarkar, G. Vozzi, Y. Tabata, J. Fiser, L. G. Zhang, 4D printing of polymeric materials for tissue and organ regeneration, *Mater. Today.* 2017, **20**(10), 577-591

[3] I. Apsite, A. Biswas, Y. Li, L. Ionov, Microfabrication using shape-transforming materials, *Adv. Funct. Mater.* 2020, 1908028

[4] Y. Hu, Z. Wang, D. Jin, C. Zhang, R. Sun, Z. Li, K. Hu, J. Ni, Z. Cai, D. Pan, X. Wang, W. Zhu, J. Li, D. Wu, L. Zhang, J. Chu. Botanical‑inspired 4D printing of hydrogel at the microscale, *Adv. Funct. Mater.* 2019, DOI:10.1002/adfm.201907377

[5] M.Y. Shie, Y.F. Shen, S. D. Astuti, A. K. Lee, S. H. Lin, N. Dwijaksara, and Y.W. Chen Review of Polymeric Materials in 4D Printing Biomedical Applications, *Polymers.* 2019, **11**(11):1864

[6] A.S. Gladman, E.A. Matsumoto, R.G. Nuzzo, L. Mahadevan, J.A. Lewis, Biomimetic 4D printing, *Nat. Mater.* 2016, **15**, 413–418

[7] H. Y. Jeong, E. Lee, S. An, Y. Lim, Y. C. Jun, 3D and 4D printing for optics and metaphotonics, *Nanophotonics*, 2020, DOI: 10.1515/nanoph-2019-0483

[8] S.Y.Zhuo, Z.G.Zhao, Z. X. Xie, Y.F. Hao, Y.C. Xu, T.Y.Zhao, H.J. Li, E.M. Knubben, L. Wen, L. Jiang, M.J. Liu, Complex multiphase organo hydrogels with programmable mechanics toward adaptive soft-matter machines *Sci. Adv.* 2020, **6**(5), eaax1464

[9] M.Wang, Y.J Chen, ,R. Khanc,H. Z. Liu, ,C. Chen,T. Chen, R.J Zhang, H. Li, A fast self-healing and conductive nanocomposite hydrogel as soft strain sensor. *Coll. Surf. A.* 2019, **567**, 139-149

[10]Y. Yamamoto, K. Kanao, T.Arie, S. Akita, K. Takei, Air Ambient-Operated pNIPAM-Based Flexible Actuators Stimulated by Human Body Temperature and Sunlight. *ACS Appl. Mater. Inter.* 2015, **7**(20),11002-11006

[11] Z. Zhao, X. Kuang, C. Yuan, H. Jerry Qi, D.N. Fang, Hydrophilic/Hydrophobic Composite Shape-Shifting Structures, *ACS Appl. Mater. Interfaces.* 2018, **10**(23), 19932-19939

[12] T.J. Wallin, J. Pikul, R.F. Shepherd, 3D printing of soft robotic systems, *Nat. Rev. Mat.* 2018, **3**, 84-100

[13]J. Koffler, W. Zhu, X. Qu, O. Platoshyn, J.N. Dulin, J. Brock, L. Graham, P. Lu, J. Sakamoto, M. Marsala, S.C. Chen, M. H. Tuszynski, Putting 3D Printing to Work to Heal Spinal Cord Injury, *Nat. Med.* 2019, **25**, 263-269.

[14] M. Malinauskas. Ultrafast laser processing of materials: from science to industry. *Light-Sci. Appl.* 2016, **5**, e16133

[15] J. Xing, M. Zheng, X. Duan, Two-photon polymerization microfabrication of hydrogels: an advanced 3D printing technology for tissue engineering and drug delivery, *Chem. Soc. Rev.* 2015, **44**(15), 5031-5039

[16] S. Kawata, H.B. Sun, T. Tanaka, K. Takada. Finer features for functional microdevices. *Nat.* 2001, **412**(6848), 697-698

[17] Y.L. Zhang, Y.Tian, H. Wang, Z.C. Ma, D.D. Han, L. G. Niu, Q.D. Chen, H.B. Sun, Dual-3D Femtosecond Laser Nanofabrication Enables Dynamic Actuation, *Acs nano*, 2019, **13**(4), 4041-4048

[18] Y. Wu, X. Hao, R. Xiao, J. Lin, Z. L. Wu, J. Yin, J. Qian, Controllable bending of bi-hydrogel strips with differential swelling. *Acta.Mech. Solida. Sin.* 2019, doi:10.1007/s10338-019-00106-6

[19] D.D.Jin, Q.Y. Chen, T.Huang, J.Y. Huang, L. Zhang, H.L.Duan, Four-dimensional direct laser writing of reconfigurable compound micromachines. *Mater.Today.* 2020, **32**, 19-25

[20] A.A. Bauhofer, Harnessing Photochemical Shrinkage in Direct Laser Writing for Shape Morphing of Polymer Sheets. *Adv. Mater.* 2017, **29**,1703024

[21] X.Ning, X. Yu, Mechanically active materials in three-dimensional mesostructures. *Sci. Adv.*2018, **4**,eaat8313

[22] D. Karalekas, A. Aggelopoulos, Study of shrinkage strains in a stereolithography cured acrylic photopolyrner resin. *J. Mater. Proc. Tech.* 2003, **136**(1-3),146-150

[23] H. L. Sun, H. A. Klok, Z. Y. Zhong, Polymers from Nature and for Nature, *Biomacromolecules*, 2018, **19**, 1697−1700

[24] Y. Zhang, X. Le, Y. Jian, L.Wei, J.Zhang, T. Chen, 3D Fluorescent Hydrogel Origami for Multistage Datan Security Protection, Adv.Funct. Mat. 2019, **29**(46):1905514

[25] A. C. Almeida, J. Canejo, S. Fernandes, C. Echeverria, P. Almeida, M. Godinho, Cellulose-Based Biomimetics and Their Applications, *Adv. Mater.* 2018, **30**, 1703655

[26] A.Urrios, C. Parra-Cabrera, N. Bhattacharjee, A.M. Gonzalez-Suarez, L.G. Rigat-Brugarolas, 3D-printing of transparent bio-microfluidic devices in PEG-DA, *Lab Chip*, 2016, **16**, 2287−2294

[27] A.Tudor, C.Delaney, H. Zhang, Fabrication of soft, stimulus-responsive structures with sub-micron resolution via two-photon polymerization of poly(ionic liquid)s, *Mater. Today*. 2018, **21**(8), 807-816

[28] L. Wolski, M. Ziolek, Insight into pathways of methylene blue degradation with $H_2O_2$ over mono and bimetallic Nb, Zn oxides. *Appl. Cata. B-Enviro.* 2018, **224**, 634-647

[29] M. A. Tasdelen, V. Kumbaraci, S. Jockusch, N.J. Turro, N. Talinli, Photoacid generation by stepwise two-photon absorption: Photoinitiated cationic polymerization of cyclohexene oxide by using benzodioxinone in the presense of iodonium salt. *Macromolecules*. 2008, **41**, 295-297

[30] Z. Moosavi-Tekyeh, N. Dastani, Intramolecular hydrogen bonding in N-salicylideneaniline: FT-IR spectrum and quantum chemical calculations. *J. Mol. Struct.* 2015, **1102**, 314-322

[31] Y. Liu, B. Shaw, M.D. Dickey, J. Genzer, Sequential self-folding of polymer sheets. *Sci. Adv.* 2017,**3**, e1602417

[32] H. Liang, L. Mahadevan, Growth, geometry, and mechanics of a blooming lily. *Proc. Natl.Acad. Sci. U.S.A.* 2011, **108**, 5516–5521

[33] J. Van Hoorick, Cross-Linkable Gelatins with Superior Mechanical Properties Through Carboxylic Acid Modification: Increasing the Two-Photon Polymerization Potential. *Biomacromolecules*. 2017, **18**(10), 3260-3272.

[34]E.D.Lemma, Mechanical Properties Tunability of Three-Dimensional Polymeric Structures in Two-Photon Lithography. *IEEE. Trans. Nanotechnol.* 2017, **16** (1) ,23-31

[35] S. Wei, W. Lu, X. Le, C. Ma, H. Lin, B. Wu, J. Zhang, P. Theato, T. Chen, Bioinspired Synergistic Fluorescence-Color Switchable Polymeric Hydrogel Actuator, *Angew. Chem. Int. Ed.* 2019, **58**(45), doi:10.1002/anie.201908437.

[36] R. C. P. Verpaalen, M. P. Cunha, T. A. P. Engels, M. G. Debije, and A. P. H. J. Schenning, Liquid Crystal Networks on Thermoplastics: Reprogrammable Photo-Responsive Actuators, *Angew. Chem. Int. Ed.* 2020, **59**, 2–7.

[37] B. Han, Y.L. Zhang, L. Zhu, Y. Li, Z.C. Ma, Y.Q. Liu, X.L. Zhang, X.W. Cao, Q.D. Chen, C.W. Qiu, H.B. Sun, plasmonic-assisted graphene oxide artificial muscles, *Adv. Mat.* 2018, **31**, 1806386

[38] R.C. Lan, J. Sun, C. Shen, R. Huang, Z.P. Zhang, L.Y. Zhang, L. Wang, H. Yang, Near-infrared photodriven self-sustained oscillation of liquid crystalline network film with predesignated polydopamine coating, *Adv. Mater.* 2020, 1906319.

[39] Y.F. Tao, C.Y.R. Wei, J. W. Liu, C.S. Deng, S.Cai, W. Xiong, Nanostructured electrically conductive hydrogels via ultrafast laser processing and self-assembly *Nanoscale*. 2019, **11**(18), 9176-9184

[40] H. Zhang, X. G. Guo, J. Wu, D. Fang, Y. H. Zhang. Soft mechanical metamaterials with unusual swelling behavior and tunable stress-strain curves. *Sci. Adv.* 2018, **4**, eaar8535

[41] N.W. Tschoegl, W.G. Knauss, I. Emri, Poisson's Ratio in Linear Viscoelasticity-A Critical Review. *Mech. Time-Depend. Mater.* 2002, **6** (1),3-51